\documentclass[twoside,english]{elsarticle}
\usepackage[T1]{fontenc}
\usepackage[latin9]{inputenc}
\usepackage{amsmath}
\usepackage{graphicx}

\makeatletter
\makeatletter
\def\ps@pprintTitle{%
   \let\@oddhead\@empty
   \let\@evenhead\@empty
   \let\@oddfoot\@empty
   \let\@evenfoot\@oddfoot
}
\makeatother

\makeatother

\usepackage{babel}
\begin{document}

\begin{frontmatter}{}

\title{Analytical Approximate Solution of a Coupled Two Frequency Hill's
Equation}

\author[VS]{Varun Saxena\corref{cor1} }

\ead{varunsaxena@mail.jnu.ac.in}

\cortext[cor1]{Corresponding author}

\address[VS]{School of Engineering, Jawaharlal Nehru University, New Mehrauli
Road, Delhi 110067, (India)}
\begin{abstract}
A coupled two frequency Hill's equation is solved. Analytically approximate
solution correct up-to first order is derived using modified Lindstedt-Poincare
perturbation method. For a wide range of controlling parameters we
compare the numerical and analytical solutions. The solution is the
first step towards developing a comprehensive understanding of the
electrodynamics of charged particles in a combinational ion trap utilizing
both electrostatic DC and RF fields along with a constant static magnetic
field with prospects of confining antimatter such as anti hydrogen
for a reasonably long durations of time. 
\end{abstract}
\begin{keyword}
Hill's equation \sep Perturbation methods\sep Modified Lindstedt
Poincare method \sep Paul trap \sep Penning trap
\end{keyword}

\end{frontmatter}{}

\section{Introduction}

Hill's equation {[}1{]} is a second order differential equation with
periodic coefficients. The equation can be described as
\begin{equation}
\ddot{x}+f(t)x=0\label{eq:hill equation}
\end{equation}
where $f(t)$ is a period function, often a combination of several
cosine and sine functions. Hill's equations finds application in several
diverse areas of applied sciences. The differential equation appears
in several settings, such as, in the analysis of lunar stability {[}2{]},
modeling of quadrupole mass spectrometer {[}3{]}, the dynamics of
an electron in a crystal using one dimensional Schrodinger equation
{[}4{]}, in a two level system in quantum optics {[}5{]} and electromagnetic
ion traps {[}6{]} in which electrostatic DC and RF fields are used
to confine charged particles in a limited space in a perturbation
free environment. 

A well known equation arising out of Eq. (\ref{eq:hill equation})
is the Mathieu equation
\begin{equation}
\ddot{x}+(a-2q_{1}\cos(2t))x=0\label{eq:mathieu}
\end{equation}
Controlling parameters, $a$, $q_{1}$ determine the stability of
the solution of Eq. (\ref{eq:mathieu}). For example, if $a=0$, the
solutions up-to $q_{1}=0.9$ are stable. Stability of Eq. (\ref{eq:mathieu})
is well documented in the literature. The equation governs the dynamics
of charged particles inside an electromagnetic ion trap, namely, Paul
trap, wherein charged particles are under the influence of electrostatic
DC and RF fields only. The coefficients $a$ and $q_{1}$ are proportional
to the applied voltage strengths. 

In this paper, we attempt to derive analytical approximate solution
for a coupled two frequency Hill's equation {[}7{]} which can be written
as 
\begin{eqnarray}
\ddot{x}-p\dot{y}+(a-2q_{1}\cos(2\eta^{-1}t)-2q_{2}\cos(2t))x & = & 0\nonumber \\
\ddot{y}+p\dot{x}+(a-2q_{1}\cos(2\eta^{-1}t)-2q_{2}\cos(2t))y & = & 0\label{eq:coupled hill}
\end{eqnarray}
Such a coupled system has recently gained importance to study the
electrodynamics of charge particles relevant to particle confinement
using two radio frequencies {[}7, 15{]} in a combinational trap utilizing
features of both Paul and Penning trap. In context to such a trapping,
coefficients $a$, $q_{1}$, $q_{2}$ are proportional to the applied
electrostatic DC and RF (radio frequency) and $p$ is the proportional
to the applied magnetic field. 

To get a better understanding of how Eq. (\ref{eq:coupled hill})
relates to the trapping of particles inside a combinational trap,
consider the quadrupole potential in a dual frequency Paul trap given
by
\begin{equation}
\Phi(x,y,z,t)=(U_{0}+V_{1}\cos(\omega_{1}t)+V_{2}\cos(\omega_{2}t))\left((x^{2}+y^{2}-2z^{2})/r_{0}^{2}\right)\label{eq:dual potential}
\end{equation}
The electric field generated by this potential is $\overrightarrow{E}(x,y,z,t)=-\overrightarrow{\nabla}\Phi(x,y,z,t)$.
Since there exists a magnetic field $\overrightarrow{B}=B_{0}\hat{k}$
due the features attributed to a Penning trap, the net force experienced
by a charged particle of charge $Q$ and mass $M$ moving with a velocity
$\overrightarrow{v}$ is given by the Lorentz force equation $\overrightarrow{F}=-Q\overrightarrow{\nabla}\Phi+Q(\overrightarrow{v}\times\overrightarrow{B})$.
If $\overrightarrow{v}=v_{x}\hat{i}+v_{y}\hat{j}+v_{z}\hat{k}$, the
components of force in the three orthogonal directions, i. e., $F_{x}$,
$F_{y}$, $F_{z}$ are given by 
\begin{equation}
F_{x}=-(U_{0}+V_{1}\cos(\omega_{1}t)+V_{2}\cos(\omega_{2}t))\left(2x/r_{0}^{2}\right)+Qv_{y}B_{0}\label{eq:coupled 1}
\end{equation}
\begin{equation}
F_{y}=-(U_{0}+V_{1}\cos(\omega_{1}t)+V_{2}\cos(\omega_{2}t))\left(2y/r_{0}^{2}\right)-Qv_{x}B_{0}\label{eq:coupled 2}
\end{equation}
\begin{equation}
F_{z}=(U_{0}+V_{1}\cos(\omega_{1}t)+V_{2}\cos(\omega_{2}t))\left(4z/r_{0}^{2}\right)
\end{equation}
Here, $U_{0}$, $V_{1,2}$ are the applied DC and RF voltages respectively,
$\omega_{1}$, $\omega_{2}$ are the primary and secondary RF frequencies,
respectively, and $r_{0}$ is the trap dimension. Upon substituting
$\omega_{2}t=2\tau$, $F_{x}=M\ddot{x}$ , $F_{y}=M\ddot{y}$, $a=8QU_{0}/Mr_{0}^{2}$,
$q_{1,2}=-4QV_{1,2}/Mr_{0}^{2}$ , $p=2QB_{0}/\omega_{2}M$, $v_{x}=\dot{x}$,
$v_{y}=\dot{y}$ and $\omega_{2}/\omega_{1}=\eta$ in Eq. (\ref{eq:coupled 1}),
Eq. (\ref{eq:coupled 2}) and rearranging the terms, one obtains the
two coupled equations given in Eq. (\ref{eq:coupled hill}). Since
$\tau$ is actually a dummy variable, without loss of generality,
it can be replaced by $t$ in the subsequent equations. The confinement
in the $x-y$ plane is through a set of coupled differential equations
given by Eq. (\ref{eq:coupled hill}), whereas, along the $z$ axis,
the trapping is on account of a combination of DC and RF voltages,
exactly like it is in a dual frequency Paul trap.

The Lorentz force due to the magnetic field acts inwards. This increases
the stability of the charged particles simultaneously being trapped
by the application of a static and a dynamic electric field in combination
with a constant magnetic field. In recent years, the trap employing
dual frequency has gained importance since it is being viewed as a
promising option to trap anti-hydrogen. In general, charged particles
with varied charge to mass ratio can be trapped effectively inside
a dual frequency Paul trap {[}7{]}. 

To produce anti-hydrogen, positron and antiproton are to be trapped
and a magnetic field is required to trap the resulting neutral particle,
anti-hydrogen. The limitation of a conventional single frequency Paul
trap in trapping two species with different charge to mass ratio is
that the weakly confined species is pushed away from the trap center
{[}8{]}. The ALPHA experiment {[}9, 10{]} and ATRAP experiment {[}11,
12{]} rely on a variation of Penning trap using static magnetic field
for their initial confinement. However it is not possible to trap
oppositely charged particles in a Penning trap on account of the presence
of only DC electric field along with a static magnetic field. Hence
a combinational trap inheriting features of both a dual frequency
Paul trap and a Penning trap holds a lot of potential in confinement
of oppositely charged species with a large charge to mass variation
and will most certainly be a significant improvement when compared
to earlier methods utilizing both electric and magnetic fields in
a conventional single frequency Paul trap {[}13,14{]}. 

Dynamics governed by the differential equations given in Eq. (\ref{eq:coupled hill})
is therefore of great interest. It offers a starting point to the
understanding of the electrodynamics that will emerge inside a combinational
trap. In Sec. \ref{sec:Solution}, we derive the time evolution of
position of the confined particle in $x$ and $y$ direction. In Sec.
\ref{sec:Comparison}, comparison of the analytical approximate solution
with the numerical solution for a wide range of control parameters
shows the robustness of the solution to depict the particle dynamics.
Sec. \ref{sec:Conclusion-and-Discussion} contains a conclusion and
a discussion on the importance of the analytical solution. 

\section{Analytical Approximate Solution \label{sec:Solution}}

We begin expressing the equations in a concise form by writing $A(t)=a-2q_{1}\cos(2\eta^{-1}t)-2q_{2}\cos(2t)$.
The coupled differential equations in Eq. (\ref{eq:coupled hill})
can now be written as 
\begin{equation}
\ddot{x}-p\dot{y}+A(t)x=0\label{eq:coupled hill1}
\end{equation}
\begin{equation}
\ddot{y}+p\dot{x}+A(t)y=0\label{eq:coupled hill2}
\end{equation}
Multiplying Eq. (\ref{eq:coupled hill1}) by imaginary $j$ and adding
to Eq. (\ref{eq:coupled hill2}) gives
\begin{equation}
\ddot{z}-jp\dot{z}+A(t)z=0\label{eq:combined coupled hill}
\end{equation}
Where $z=y+jx$. Let $z=w(t)\exp(jpt/2)$. The function $w(t)$ is
actually a complex function which can further be substituted as $w=X+jY$.
Hence upon substituting $z=(X+jY)\exp(jpt/2)$ and after some basic
manipulations, Eq. (\ref{eq:combined coupled hill}) can be written
as 
\begin{equation}
\ddot{w}+\left(A(t)+p^{2}/4\right)w=0\label{eq:modified coupled hill}
\end{equation}
Where, $\ddot{w}=\ddot{X}+j\ddot{Y}$. Writing $a1=a+p^{2}/4$, $q_{2}=q_{r}q_{1}$,
$\Omega_{1}=2\eta^{-1}$, $\Omega_{2}=2$, Eq. (\ref{eq:modified coupled hill})
can be expressed as 
\begin{equation}
\ddot{w}+(a1-2q_{1}\cos(\Omega_{1}t)-2q_{r}q_{1}\cos(\eta\Omega_{1}t))w=0\label{eq:altered couple hill}
\end{equation}
Applying Modified Lindstedt-Poincare method {[}16{]} in Eq. (\ref{eq:altered couple hill}),
we begin by writing
\begin{eqnarray}
a1 & = & \nu^{2}+q_{1}\alpha_{1}+q_{1}^{2}\alpha_{2}\nonumber \\
x & = & x_{0}+q_{1}x_{1}+q_{1}^{2}x_{2}\label{eq:mlp}
\end{eqnarray}
Substituting the values of $a1$ and $x$ from Eq. (\ref{eq:mlp})
in Eq. (\ref{eq:altered couple hill}) and solving equations, one
at a time for $\mathcal{O}(q_{1}^{0})$, $\mathcal{O}(q_{1}^{1})$,
$\mathcal{O}(q_{1}^{2})$, we get
\begin{align}
X & =D_{1}\phi(t)+E_{1}\psi(t)\label{eq:mlp1}\\
Y & =D_{2}\phi(t)+E_{2}\psi(t)\label{eq:mlp2}
\end{align}
Where $D_{1,2}$ and $E_{1,2}$, are real constants that depend on
the initial positions and velocities of the charged particle, i.e.,
$x_{0}$, $y_{0}$, $v_{x0}$, $v_{y0}$. Moreover, one can express
$\phi(t)$ and $\psi(t)$, correct up-to first order as 
\begin{equation}
\phi(t)=\cos(\nu t)+a_{1}\cos(\nu-\Omega_{1})t+a_{2}\cos(\nu+\Omega_{1})t+a_{3}\cos(\nu-\eta\Omega_{1})t+a_{4}\cos(\nu+\eta\Omega_{1})t\label{eq:phi}
\end{equation}
\begin{equation}
\psi(t)=\sin(\nu t)+a_{1}\sin(\nu-\Omega_{1})t+a_{2}\sin(\nu+\Omega_{1})t+a_{3}\sin(\nu-\eta\Omega_{1})t+a_{4}\sin(\nu+\eta\Omega_{1})t\label{eq:psi}
\end{equation}
Where $a_{1}=q_{1}/(\nu^{2}-(\nu-\Omega_{1})^{2})$, $a_{2}=q_{1}/(\nu^{2}-(\nu+\Omega_{1})^{2})$,
$a_{3}=q_{r}q_{1}/(\nu^{2}-(\nu-\eta\Omega_{1})^{2})$, $a_{4}=q_{r}q_{1}/(\nu^{2}-(\nu+\eta\Omega_{1})^{2})$
and $\nu$ is the slow frequency given by 
\begin{equation}
\nu=\sqrt{\left[\left(a+p^{2}/4\right)+(2q_{1}^{2}/\Omega_{1}^{2})\left(1+q_{r}^{2}/\eta^{2}\right)\right]}\label{eq:slow frequency}
\end{equation}
In Eq. (\ref{eq:mlp}), the constants $a1$ and $x$ are written up-to
second order, even though independent solutions of Eq. (\ref{eq:phi}),
Eq. (\ref{eq:psi}) are written up-to first order. This has been done
to evaluate the slow frequency by deriving expressions for $\alpha_{1}$
and $\alpha_{2}$. The value of $\alpha_{1}$, to eliminate secular
terms for $\mathcal{O}(q_{1}^{1})$ comes out to be $\alpha_{1}=0$.
Similarly, the value of $\alpha_{2}$, to eliminate secular terms
for $\mathcal{O}(q_{1}^{2})$ comes out to be $\alpha_{2}=(-2/\varOmega_{1}^{2})(1+q_{r}^{2}/\eta^{2})$.
Backtracking from $X$and $Y$, the time evolution of position $x(t)$
and $y(t)$ for the charged particle is 
\begin{equation}
x=Y\cos(pt/2)+X\sin(pt/2)\label{eq:final_x}
\end{equation}
\begin{equation}
y=X\cos(pt/2)-Y\sin(pt/2)\label{eq:final_y}
\end{equation}
Its worth observing that $\dot{\phi_{0}}=\dot{\phi}(t=0)=0$ and $\psi_{0}=\psi(t=0)=0$.
If one writes $\phi_{0}=\phi(t=0)$ and $\dot{\psi}_{0}=\dot{\psi}(t=0)$,
the values of constants $D_{1,2}$ and $E_{1,2}$ come out to be,
$D_{1}=y_{0}/\phi_{0}$, $D_{2}=x_{0}/\phi_{0}$, $E_{1}=(v_{y0}+px_{0}/2)/\dot{\psi}_{0}$
and $E_{2}=(v_{x0}-py_{0}/2)/\dot{\psi_{0}}$. 

\section{Comparison of Analytical Solution with Numerical Solution \label{sec:Comparison} }

The solutions are obtained by varying the controlling parameters,
namely, $p$, $q_{1}$, $q_{2}$ and $\eta$. In Fig. 1, a comparison
of the numerical and analytical solution is shown with parameter values
$p=0.3$, $q1=0.0011$, $\eta=45$ in sub figures (a) $q2=0.15$,
(b) $q2=0.2$, (c) $q3=0.24$ and with parameter values $p=0.7$,
$q1=0.002$, $\eta=45$ in sub figures (d) $q2=0.19$, (e) $q2=0.23$,
(f) $q2=0.27$. In Fig. 2, a comparison of the numerical and analytical
solution is shown with parameter values $p=0.9$, $q1=0.002$, $\eta=45$
in sub figures (a) $q2=0.15$, (b) $q2=0.17$, (c) $q3=0.2$ and with
parameter values $p=0.3$, $q1=0.0011$, $q2=0.2$ in sub figures
(d) $\eta=5$, (e) $\eta=35$, (f) $\eta=55$. The values of $q1$
and $q2$ are proportional to the applied RF voltages $V_{1}$ and
$V_{2}$ respectively, $p$ is proportional to the applied magnetic
field strength $B_{0}$ and $\eta$ is the ratio of the secondary
voltage frequency $\omega_{2}$ and primary voltage frequency $\omega_{1}$. 

\begin{figure}
(a)\includegraphics[width=2.5in,height=2in]{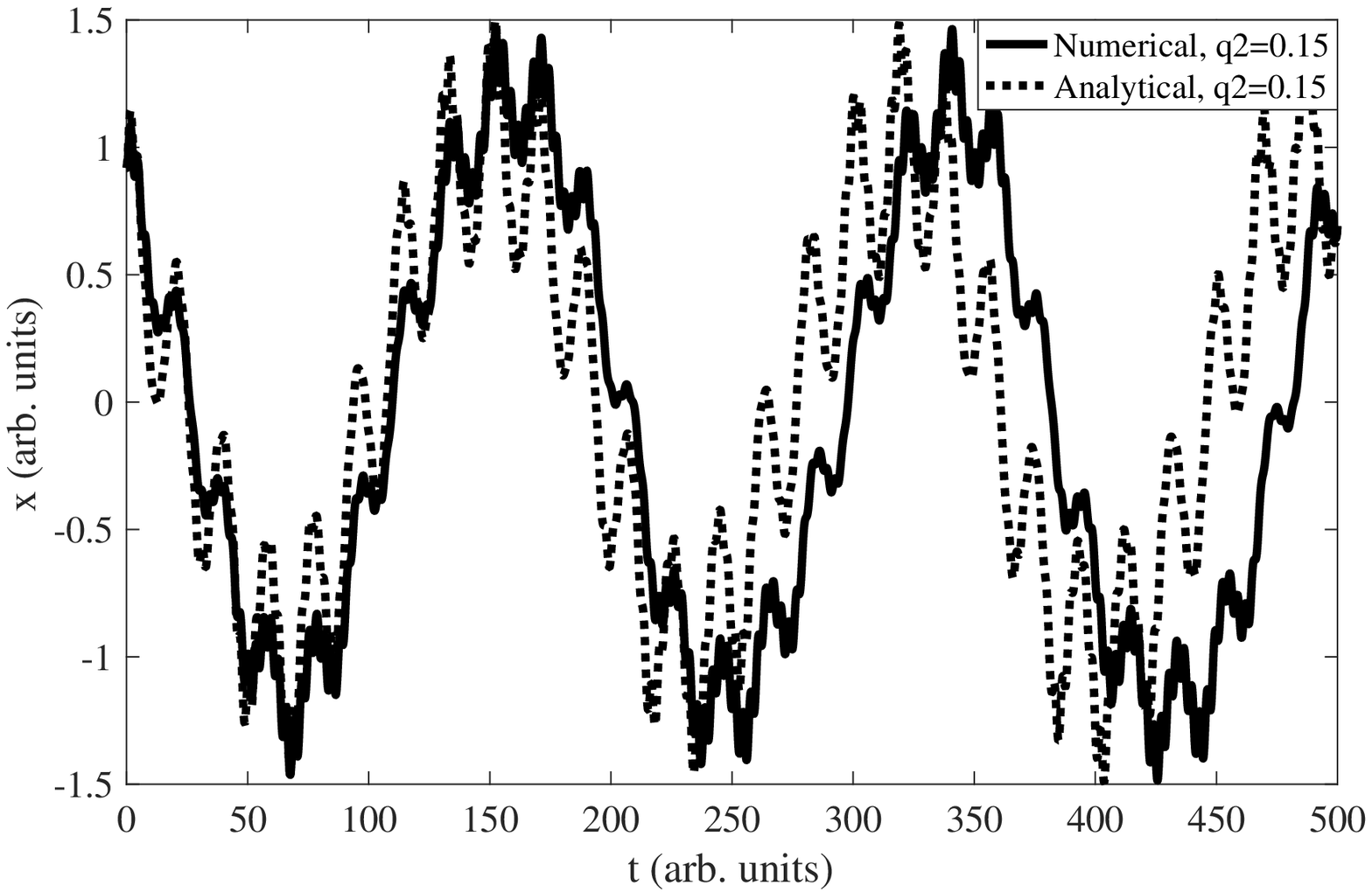}(d)\includegraphics[width=2.5in,height=2in]{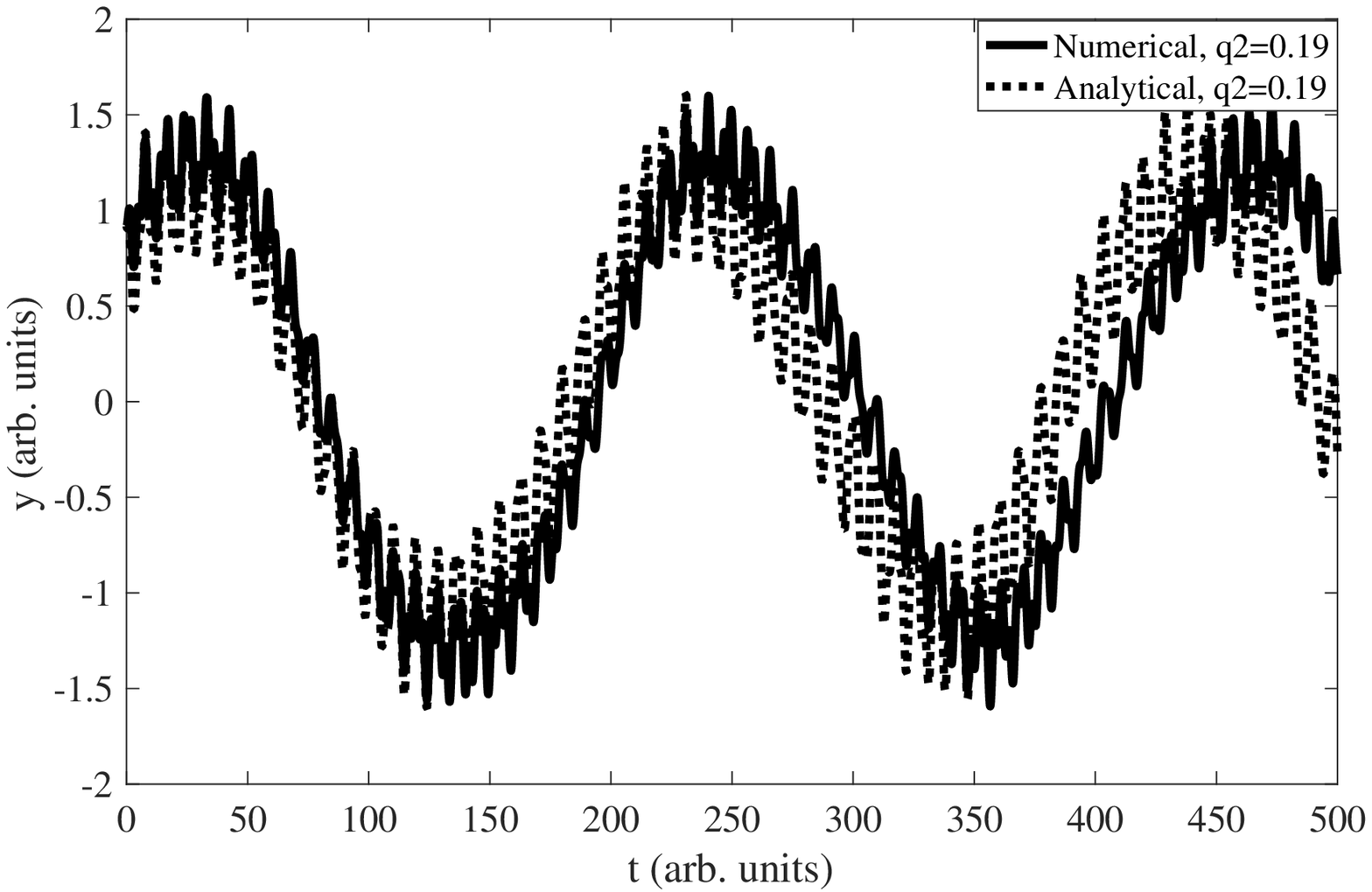}

(b)\includegraphics[width=2.5in,height=2in]{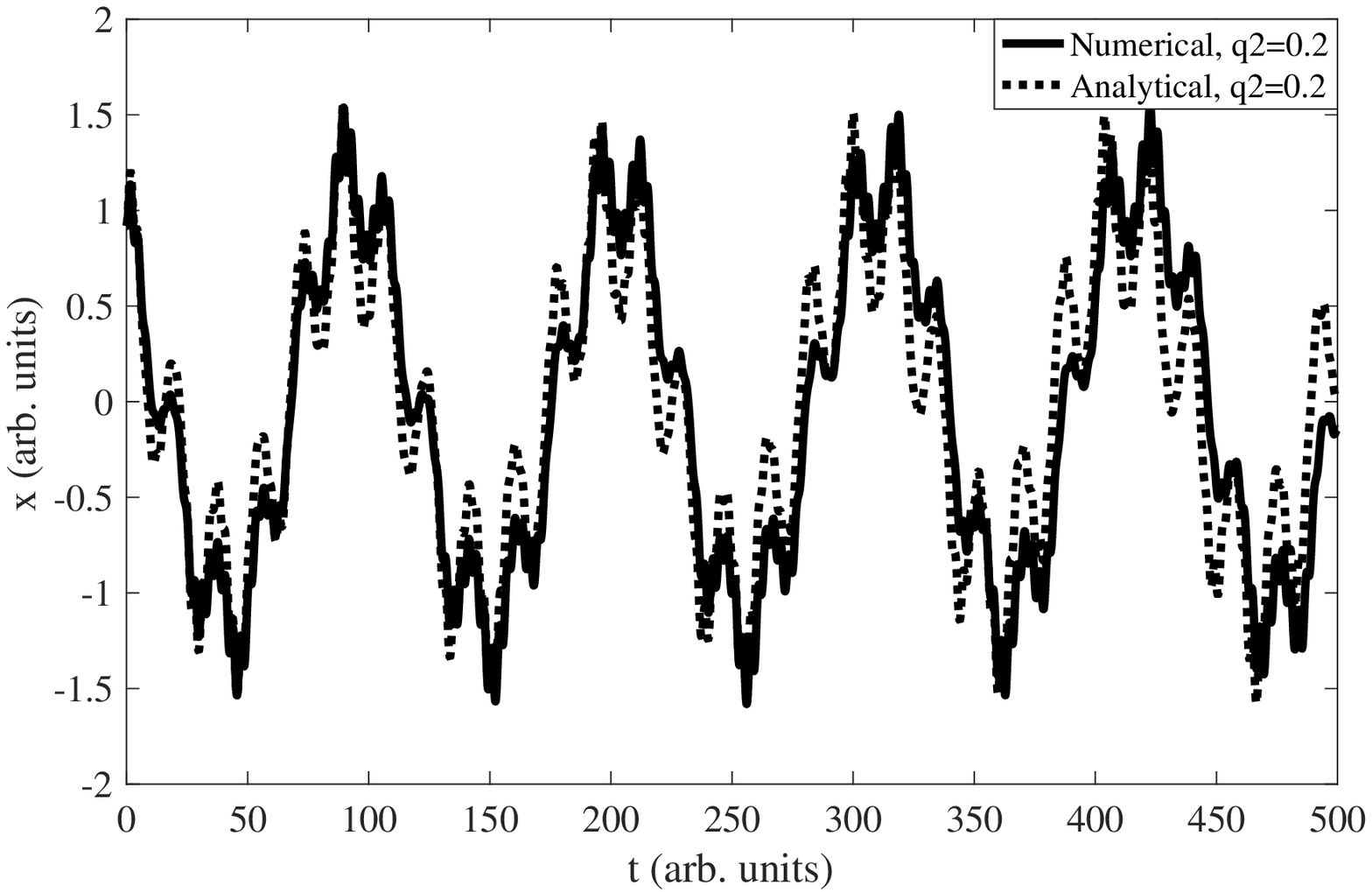}(e)\includegraphics[width=2.5in,height=2in]{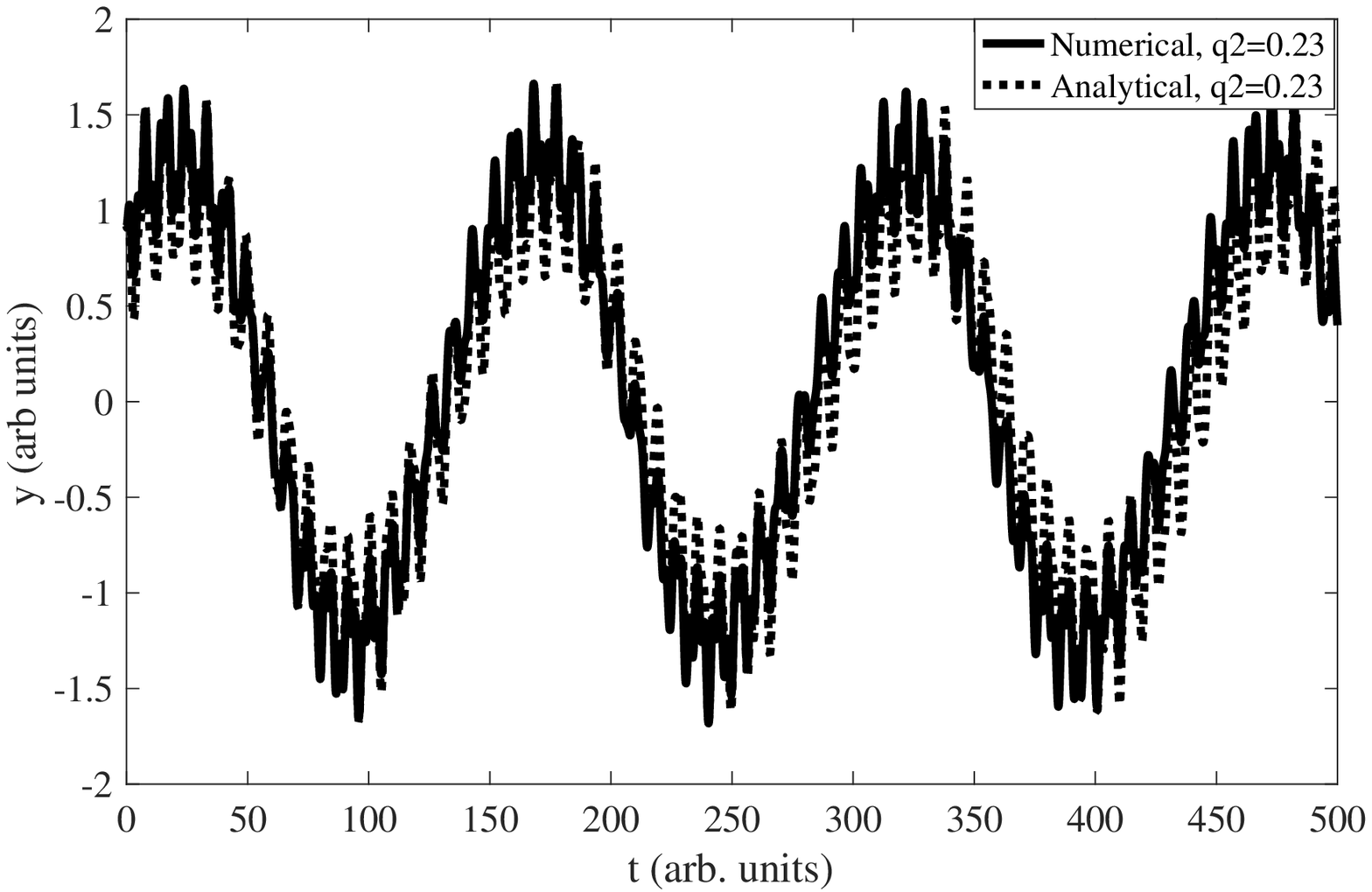}

(c)\includegraphics[width=2.5in,height=2in]{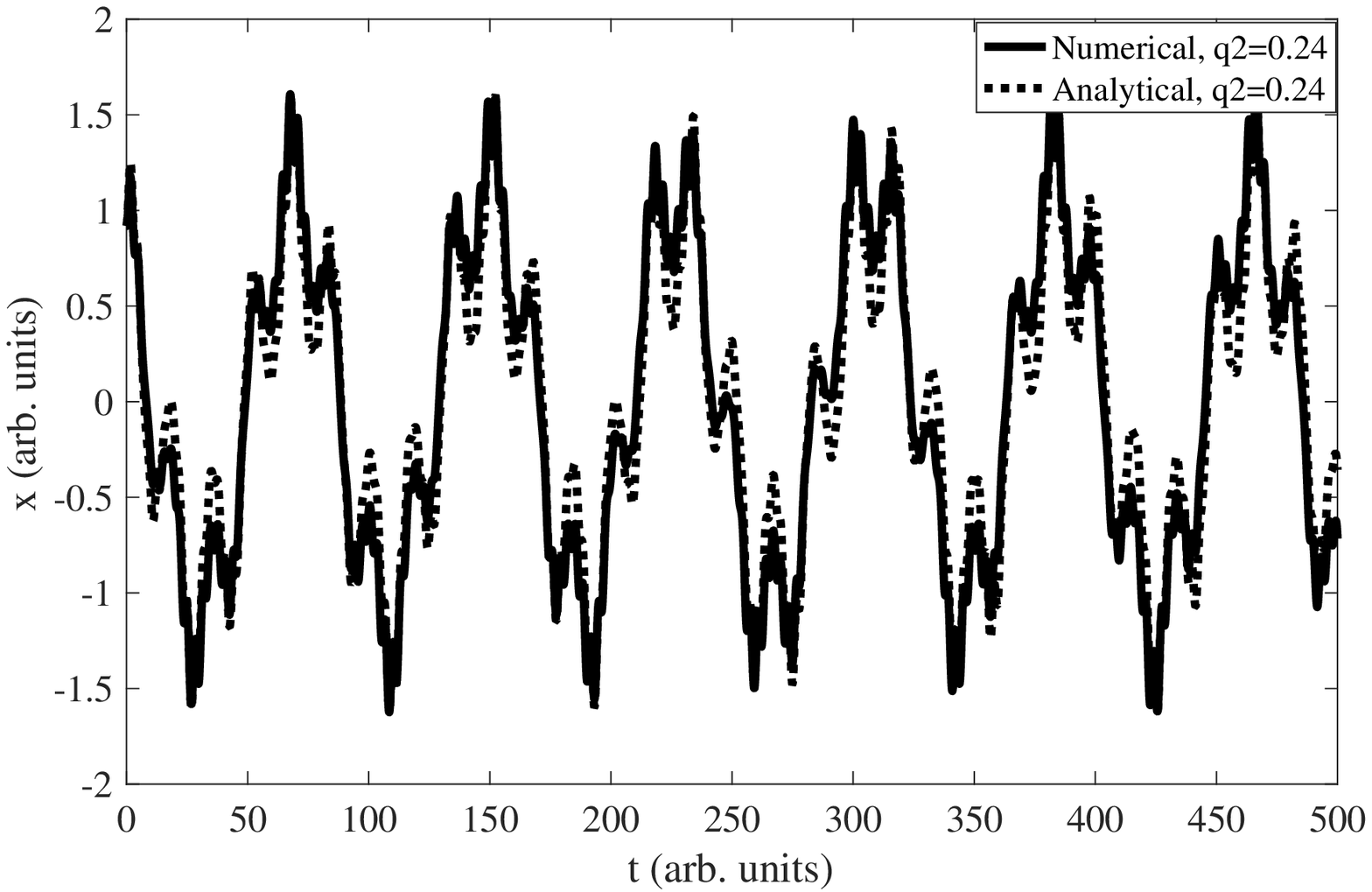}(f)\includegraphics[width=2.5in,height=2in]{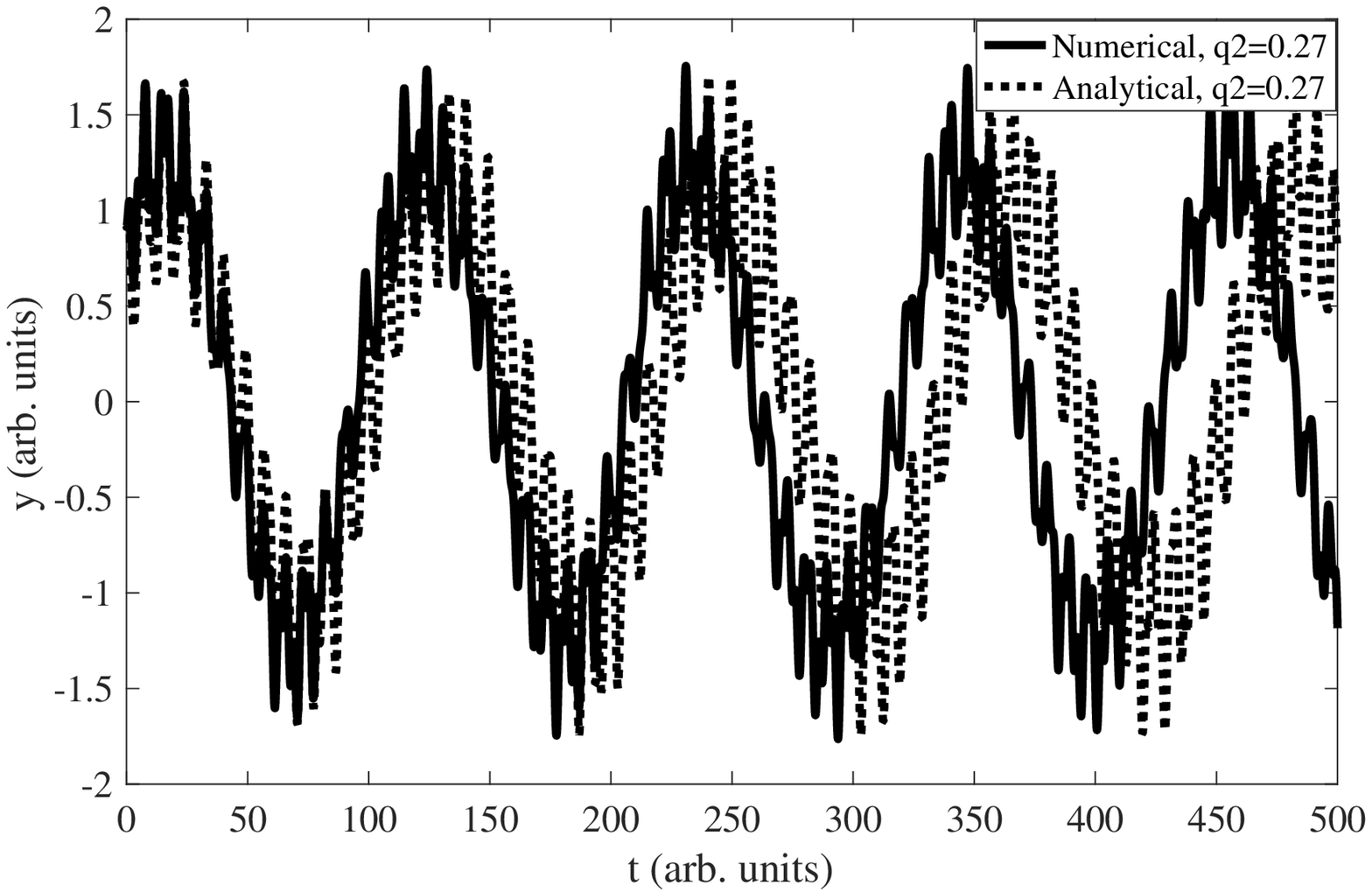}

\medskip{}

Figure 1: The plots (a), (b), (c), show a comparison of the numerical
and the analytical solutions for parameter values $p=0.3$, $q1=0.0011$,
$\eta=45$. Plots (d), (e), (f) show a comparison of the numerical
and the analytical solutions for $p=0.7$, $q1=0.002$, $\eta=45$.\label{Figure-1:-fig1}
\end{figure}

\begin{figure}
(a)\includegraphics[width=2.5in,height=2in]{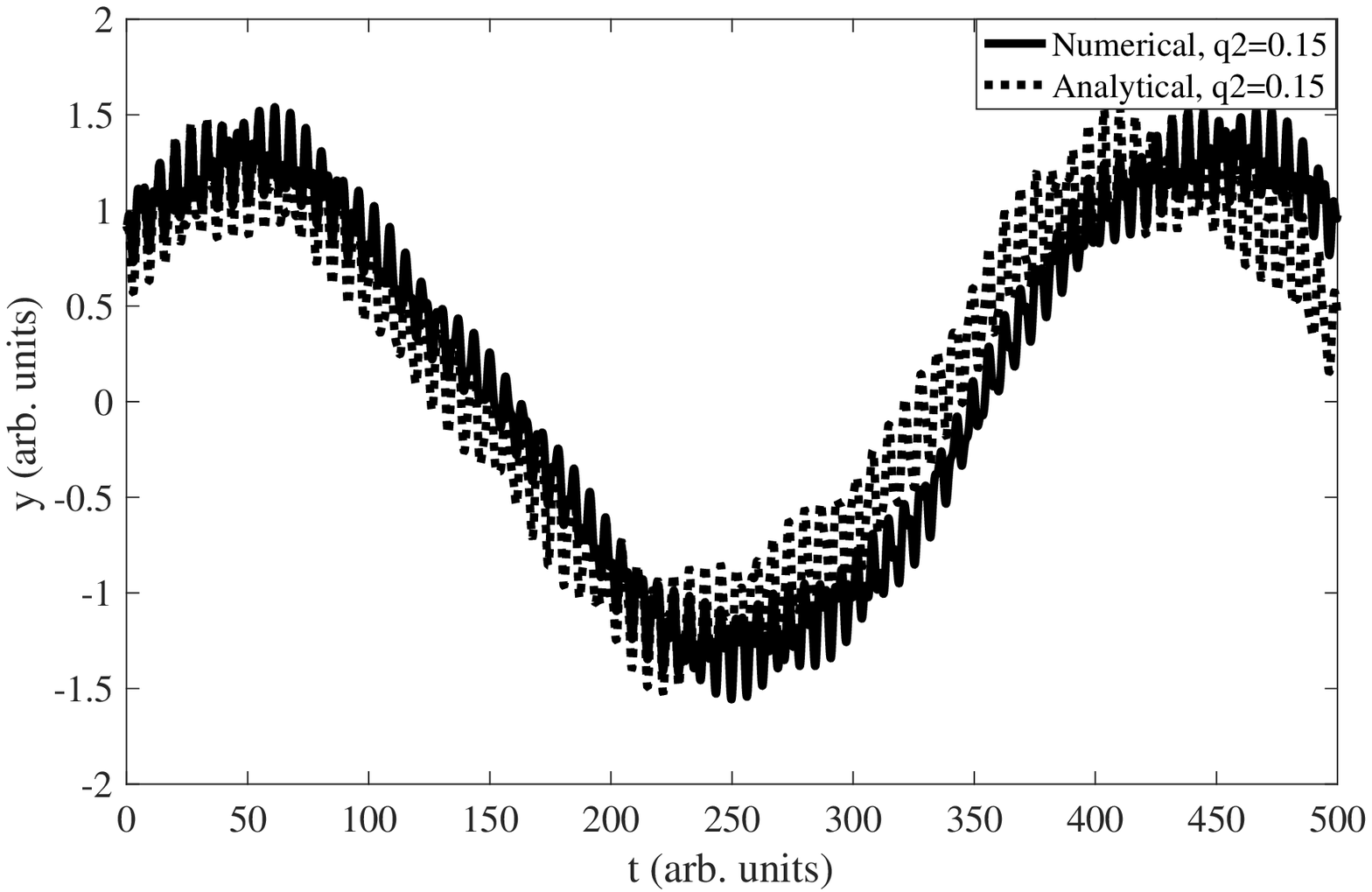}(d)\includegraphics[width=2.5in,height=2in]{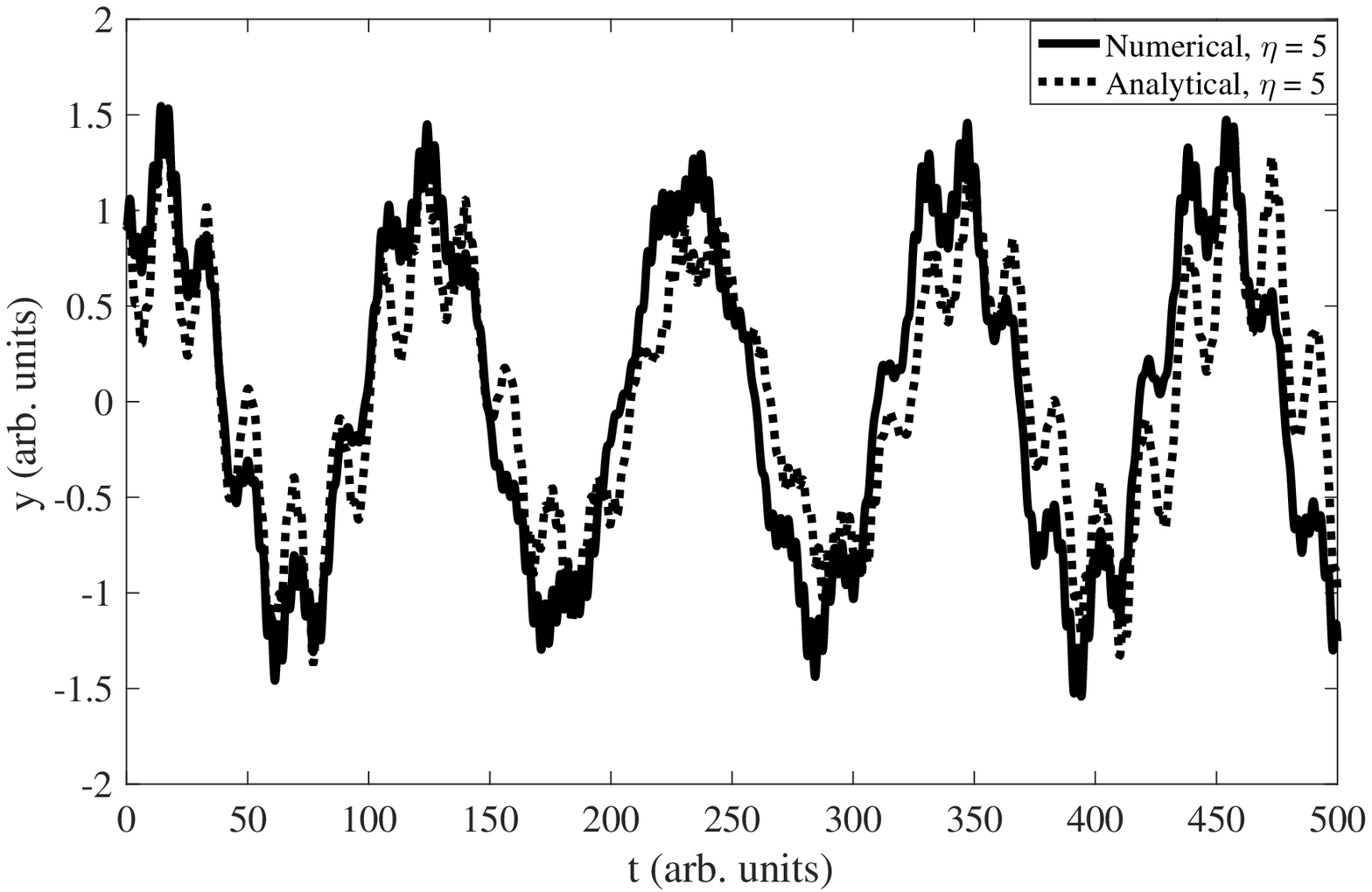}

(b)\includegraphics[width=2.5in,height=2in]{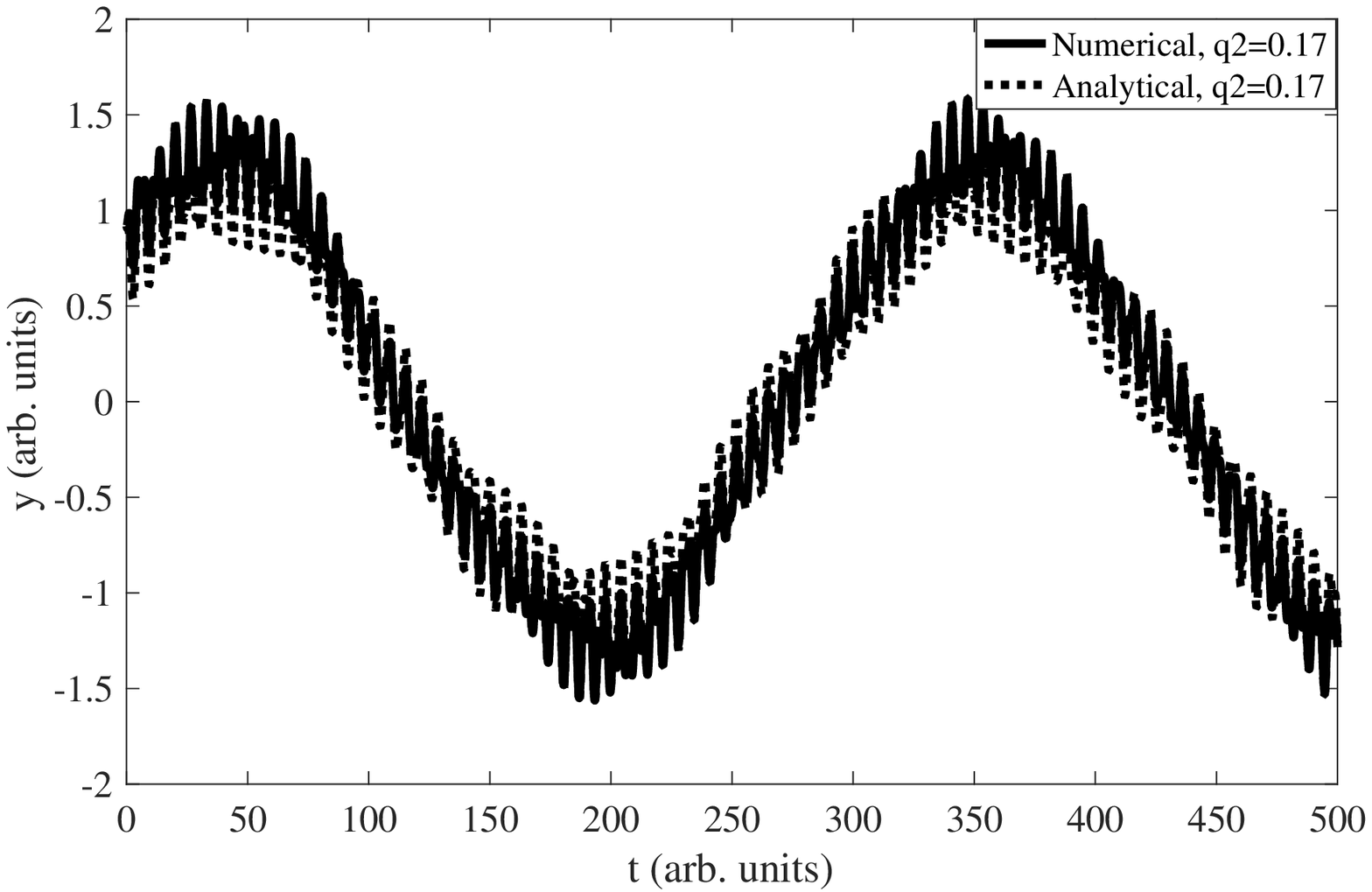}(e)\includegraphics[width=2.5in,height=2in]{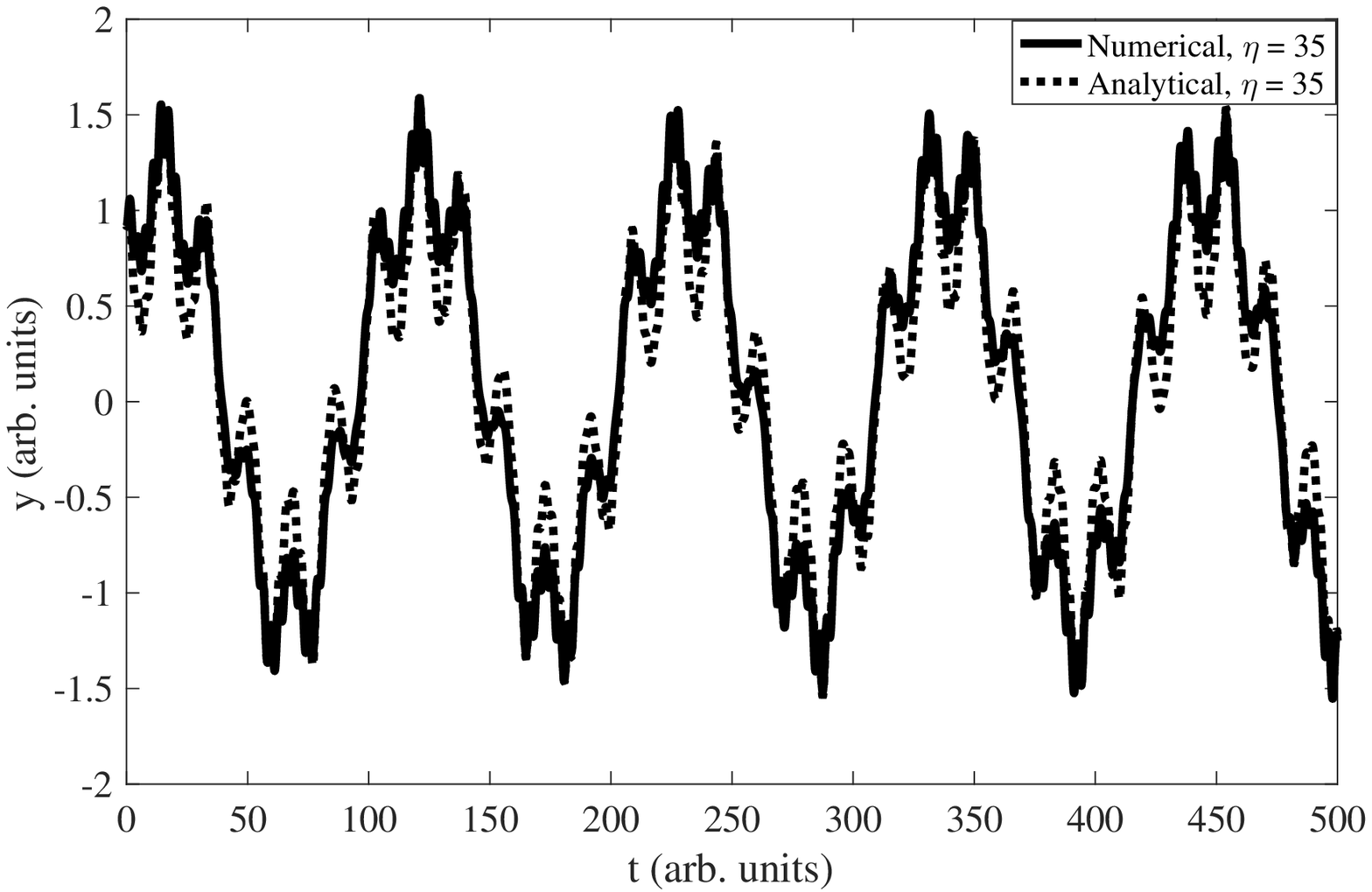}

(c)\includegraphics[width=2.5in,height=2in]{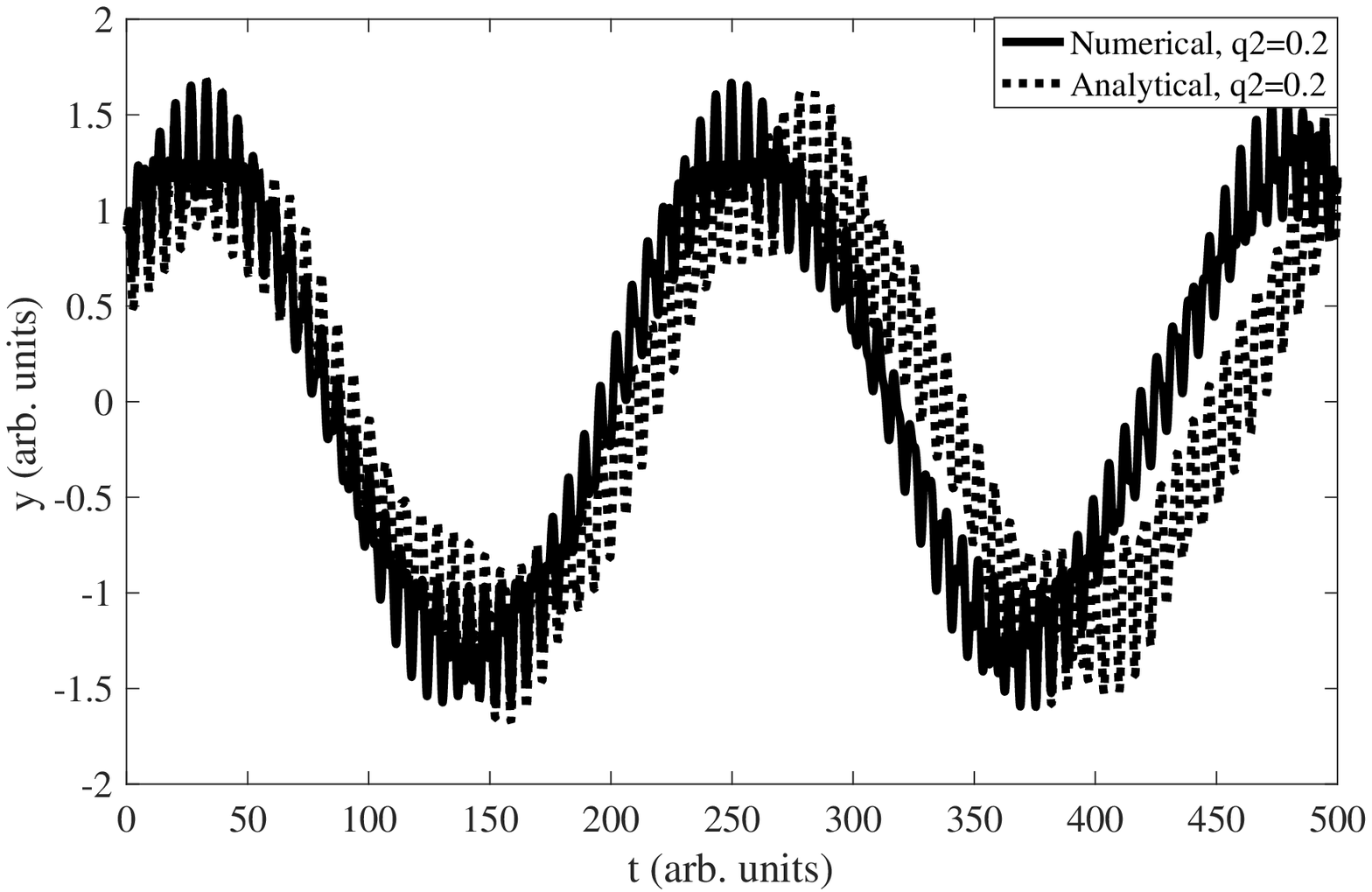}(f)\includegraphics[width=2.5in,height=2in]{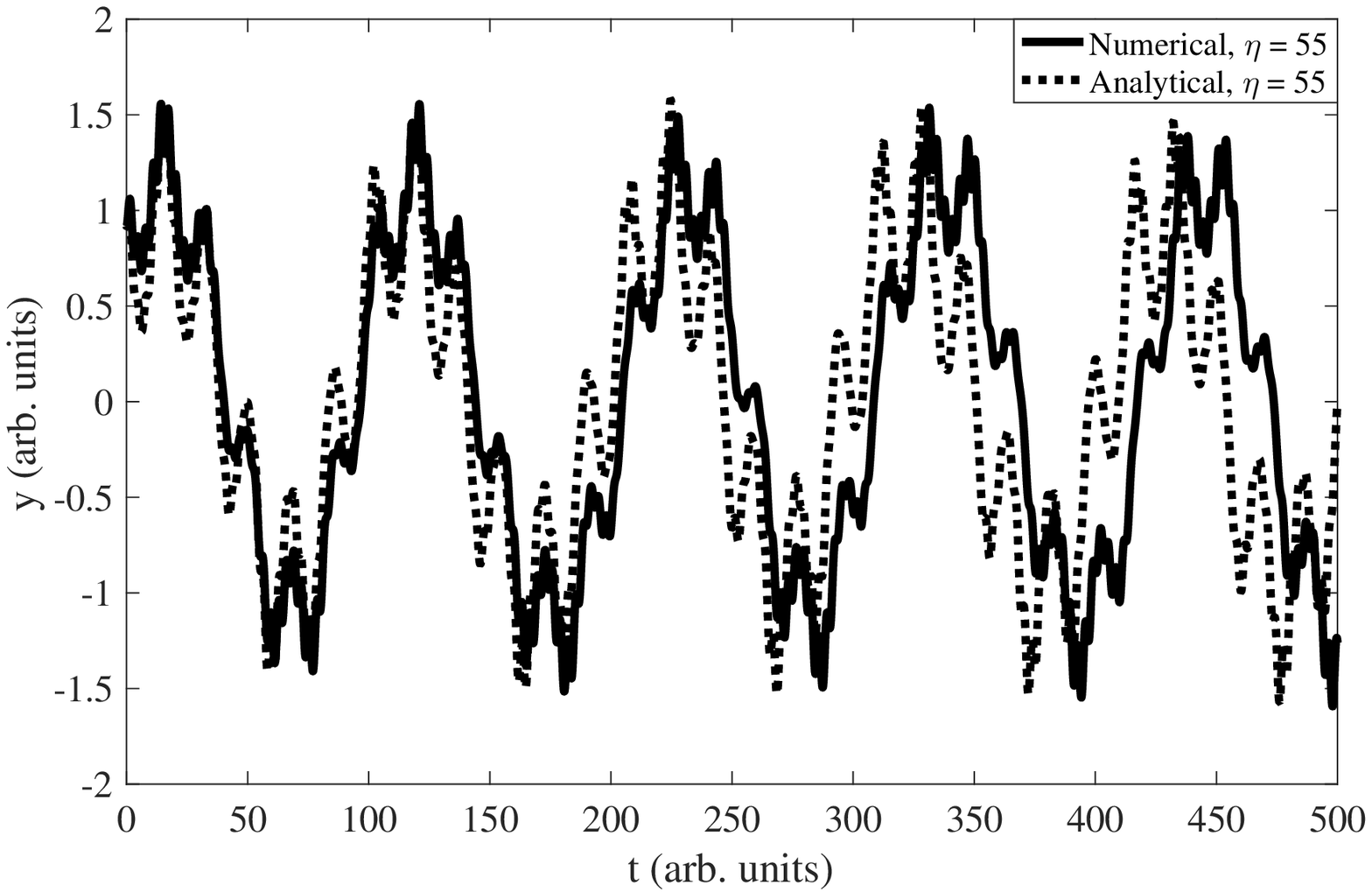}

\medskip{}

Figure 2: The plots (a), (b), (c), show a comparison of the numerical
and the analytical solutions for parameter values $p=0.9$, $q1=0.002$,
$\eta=45$. Plots (d), (e), (f) show a comparison of the numerical
and the analytical solutions for $p=0.3$, $q1=0.0011$, $q2=0.2$.\label{Figure-2:-fig2}
\end{figure}

\section{Conclusion and Discussion\label{sec:Conclusion-and-Discussion}}

For particle trajectory in $x-y$ plane, the analytical approximate
solution correct up-to first order, is derived for the coupled two
frequency Hill's equation using modified Lindstedt-Poincare method.
The analytical solution matches well with the numerical solution obtained
by numerical simulating the system of coupled differential equations
given in Eq. (3). The analytical solution has a limited number of
harmonic terms, i.e., $(\nu\pm\Omega_{1})$ and $(\nu\pm\eta\Omega_{1})$
terms, whereas the numerical solution encompasses the effect of all
the harmonic terms which make up the complete solution. Therefore,
the matching is observed for some range of controlling parameters
only. If the order of the analytical solution is increased, the range
of operating parameters for which the two solutions match will widen.
However, the derivation of such higher order terms will be mathematically
challenging. It is important to see that the solution described by
Eq. \ref{eq:phi} and Eq. \ref{eq:psi} will blow up when $\eta\sim1$.
To obtain single frequency solutions one can simply substitute $q_{r}=0$
and keep $\eta$ away from $1$. In most of the practical settings
{[}7,15{]}, the value of $\eta$ is substantially higher than $1$,
a regime wherein the analytical solutions are a good match to the
numerical solutions. 

Experience guides us that analytical solution correct up-to first
and second order are usually sufficient to provide deeper insights
to both individual particle as well as collective dynamics inside
the trap {[}17-19{]}. The relevance of an analytical solution cannot
be understated when one has to study the collective dynamics inside
such combinational traps. Since the fields are spatially linear in
this set up, one has to see if a distribution function can be constructed
for the particles by the method of inversion {[}17{]}. It is well
known that RF heating on account of applied RF fields will increase
the temperature of the charged particles. The analytical tracking
of temperature variation for each species inside such a trap is therefore
important {[}18-19{]}. Temperature can be evaluated as the second
order moment of the distribution function. To the best of my knowledge,
such analytical work on collective dynamics for combinational traps
has not been undertaken. Going ahead in this direction will require
us to choose some operating parameters for stable configuration. The
analytical expressions for particle dynamics derived in this work
assumes importance as a vital starting point.

Imperfections in electrode geometry of the trap introduce deviations
from the quadrupole potential. It would be interesting to see if analytical
solutions can be derived for particle dynamics in such a scenario.
Study of non linear resonances, deviation in the values of secular
frequencies, changes in the stability regimes of the dynamics are
all very interesting problems that could be taken up as future work.

\end{document}